\documentclass[aps,reprint,superscriptaddress,nofootinbib]{revtex4-1}
\usepackage{bm,amsmath,amssymb}
\usepackage{graphicx}
\usepackage{color}
\usepackage{txfonts}
\usepackage[colorlinks=true,urlcolor=blue,citecolor=blue,linkcolor=blue,breaklinks=true]{hyperref}

\input glyphtounicode
\graphicspath{{figs/}}

\begin{document}

\title{Probing and controlling spin chirality in Mott insulators by circularly polarized laser}

\author{Sota Kitamura}
\affiliation{Department of Physics, University of Tokyo, Hongo, Tokyo 113-0033, Japan}

\author{Takashi Oka}
\affiliation{Max-Planck-Institut f{\"u}r Physik komplexer Systeme, N{\"o}thnitzer Stra{\ss}e 38, 01187 Dresden, Germany}
\affiliation{Max-Planck-Institut f{\"u}r Chemische Physik fester Stoffe, N{\"o}thnitzer Stra{\ss}e 40, 01187 Dresden, Germany}

\author{Hideo Aoki}
\affiliation{Department of Physics, University of Tokyo, Hongo, Tokyo 113-0033, Japan}
\affiliation{Electronics and Photonics Research Institute, Advanced Industrial Science and Technology (AIST), Tsukuba, Ibaraki 305-8568, Japan}
\begin{abstract}

Scalar spin chirality, a three-body spin correlation 
that breaks time-reversal symmetry, is revealed to couple directly to circularly polarized laser. 
This is shown by the Floquet formalism 
for the periodically driven repulsive Hubbard model with a strong-coupling expansion.  
A systematic derivation of the effective low-energy Hamiltonian
for a spin degree of freedom reveals that 
the coupling constant for scalar spin chirality 
can become significant for a situation in which the driving frequency
and the on-site interaction are comparable.  
This implies that the scalar chirality can be induced by circularly polarized lights, or that
it can be used conversely for \textit{probing} the chirality in Mott insulators as a circular dichroism.
\end{abstract}

\date{\today}

\maketitle

\section{Introduction}
While magnetism is a fundamental manifestation of electron correlation
as typified by the coupling between spins emerging from 
the kinetic exchange processes for Mott-insulating electrons in 
the Hubbard model, such a link between charge
and spin degrees of freedom also yields characteristic electromagnetic
responses peculiar to Mott insulators~\cite{Bulaevskii2008}.

This has led us in the present work to pose a question: 
can illuminating a strongly-correlated electron system by a laser control spin structures?  
Indeed, periodically driven systems have recently become an important platform for novel nonequilibrium phenomena.  
Theoretically, the Floquet formalism dictates that 
a periodically driven system can be mapped 
onto an effective static Hamiltonian. The effective Hamiltonian can accommodate new phases of matter  
that would be unimaginable in equilibrium, where a prime example of 
Floquet engineering is the ``Floquet topological insulator"~\cite{Oka2009,Kitagawa2011,Lindner2011}. 
It was initially proposed for 
two-dimensional Dirac systems irradiated by a circularly polarized
laser. There, 
the effective Hamiltonian turned out to be the Haldane model~\cite{Haldane1988} 
for the anomalous quantum Hall effect in a honeycomb lattice, which, with imaginary hopping amplitudes, 
has been thought to be quite an unrealistic toy model. 
Derivation of the Haldane model in the Floquet formalism was done in Ref.~\cite{Kitagawa2011}, 
for driving frequencies higher than the electronic band width,
with a perturbative expansion from the infinite driving
frequency (the high-frequency expansion).

These works arouse interest in the properties of periodically driven
Mott insulators. 
For instance, Ref.~\cite{Mikami2016} discusses these 
and reveals transitions between Floquet topological insulators and Mott insulators, 
but magnetic correlations are not explicitly dealt with there.  
An interesting possibility then is that the spin degree of freedom, too, would be controlled
by periodic electric fields in Mott insulators, 
since the coupling between spins is derived from the electron correlation~\cite{Mentink2014,Bukov2016}. 
Indeed, a seminal work~\cite{Mentink2014} shows that 
the Heisenberg exchange interaction $J$ is modulated, 
even from antiferromagnetic to ferromagnetic,
in strong linearly polarized lasers. 
The magnetic component of a laser is also shown to induce magnetic orders
in spin systems~\cite{Takayoshi2014,Takayoshi2014-2},
but here we concentrate on the effects of electric fields. 
Thus a specific question we pose in the present paper is as follows: 
when a Mott insulator is illuminated by a circularly polarized laser, 
what kind of new spin states can emerge, 
especially topological states from an interplay of strong correlation and circularly polarized light?

\begin{figure}
\begin{centering}
\includegraphics{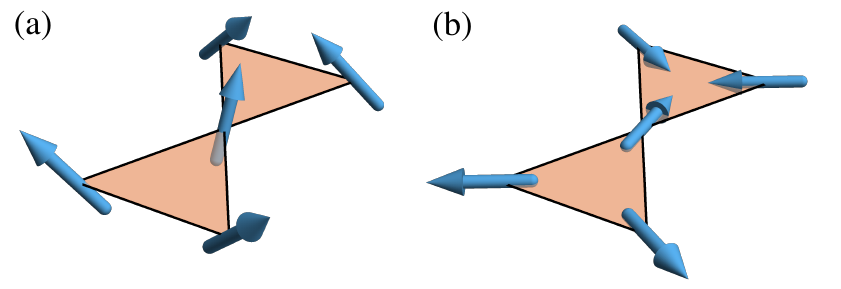}
\caption{(a) A schematic spin configuration with a nonzero scalar spin chirality 
$(\bm{S}_i\times\bm{S}_j)\cdot\bm{S}_k$.  
Note that the scalar spin chirality can be nonzero without breaking the SU(2) symmetry.
Also, while the scalar spin chirality is depicted on a triangular 
cofiguration here, the chirality revealed in this paper emerges on 
nonfrustrated lattices (e.g., square) as well when periodically driven. 
(b) A spin configuration with a nonzero vector spin chirality
$\sum\bm{S}_i\times\bm{S}_j$.
\label{fig:chirality}}
\end{centering}
\end{figure}

We expect the emergence of scalar spin chirality, which is spin correlations 
$(\hat{\bm{S}}_{i}\times\hat{\bm{S}}_{j})\cdot\hat{\bm{S}}_{k}$ 
forming finite solid angles as illustrated in Fig.~\ref{fig:chirality}(a), 
when a Mott insulator is irradiated by a circularly polarized laser.
This can be readily seen in the high-frequency limit, 
since the leading-order effective Hamiltonian should then be the Hubbard model
with complex hopping amplitudes as in the Haldane model. 
The scalar spin-chirality term is indeed known to appear in the third order of 
the strong-coupling expansion if the hopping amplitude is complex~\cite{Sen1995}.

\begin{figure}
\begin{centering}
\includegraphics{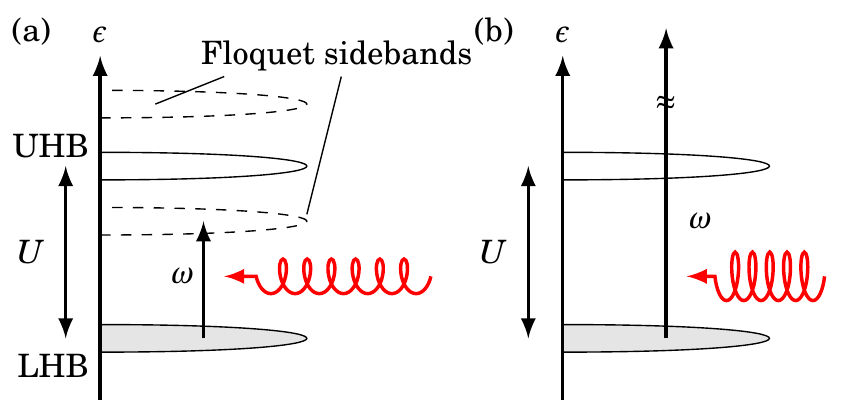}
\caption{Systems considered in the present study.
(a) The situation we are mainly interested in,
where the driving frequency is comparable but off-resonant with the charge gap.
UHB and LHB denote the upper and lower Hubbard bands, respectively.
(b) The situation where the conventional high-frequency expansion is applicable.
\label{fig:low-freq-driving}}
\end{centering}
\end{figure}

However, the high-frequency expansion is justified 
only when the driving frequency is higher than 
all the relevant energy scales in the system, 
such as the electronic band width and the on-site interaction [see Fig.~\ref{fig:low-freq-driving}(b)]. 
As the on-site repulsion in Mott insulators is typically a few eV, 
the applicable range of the high-frequency expansion becomes quite limited in realistic situations. 
So we need to extend the applicability of the emergent
spin chirality to the case in which the driving frequency is smaller 
than the charge gap arising from the on-site repulsion [see Fig.~\ref{fig:low-freq-driving}(a)].
In the first part of the present paper, we have done this 
by systematically formulating the strong-coupling expansion 
for the periodically driven Hubbard model. 
We derive an effective spin Hamiltonian that has time-periodic couplings 
by eliminating the charge-excitation degree of freedom 
when the driving frequency does not cause 
a resonance between the upper and lower Hubbard bands. 
We then perform an expansion, 
that is valid when the driving frequency is higher than the exchange interaction 
even when the frequency is smaller than the charge gap. 
Intriguingly, the chiral coupling term obtained in this manner reveals that 
the coupling is actually much larger for $\omega\sim U$ than for $\omega\gg U$.

If scalar spin chirality exists, one remarkable consequence is 
the emergence of the ``chiral spin liquid" phase. 
This phase is investigated for a complex Hubbard model and 
the derived spin Hamiltonian for static systems~\cite{He2011,Bauer2014,Hickey2016},
and even for a periodically driven system in a recent study~\cite{Claassen2016}.
We shall show that the scalar chiral coupling induced by circularly polarized lights 
can exceed a critical value estimated for the chiral spin liquid.  
Another point we shall note is that 
the emergent scalar-chirality term has a contribution 
proportional to the driving amplitude squared. 
This implies that the presence of scalar chirality 
in the Mott insulator should be fingerprinted in the dielectric function. 
These findings are expected to be useful for understanding and controlling characteristic materials 
such as herbertsmithite ZnCu$_{3}$(OH)$_{6}$Cl$_{2}$~\cite{Norman2016},
which is an antiferromagnet on a frustrated lattice and known to be a spin liquid,
and PdCrO$_2$~\cite{Takatsu2014},
which is suggested to have finite scalar spin chirality.

\section{Effective low-energy Hamiltonian}

Let us start with the formulation. We consider
the periodically driven Hubbard model, 
\begin{equation}
\hat{H}_{\text{Hub}}(t)=-\sum_{ij\sigma}t_{ij}(t)\hat{c}_{i\sigma}^{\dagger}\hat{c}_{j\sigma}
+\frac{1}{2}U\sum_{i}\hat{n}_{i}(\hat{n}_{i}-1)\label{eq:td-hubbard}
\end{equation}
at half filling, where $U$ is the on-site repulsion with 
$\hat{n}_{i}=\sum_{\sigma}\hat{c}_{i\sigma}^{\dagger}\hat{c}_{i\sigma}$. 
The time-dependent hopping amplitude $t_{ij}(t)$ is represented by
the Peierls substitution as
\begin{equation}
t_{ij}(t)=t_{ij}e^{-i\bm{A}(t)\cdot\bm{R}_{ij}},
\end{equation}
where $t_{ij}$ is the bare hopping amplitude,
$\bm{A}(t)=(1/2)(\bm{A}e^{-i\omega t}+\bm{A}^{\ast}e^{i\omega t})$
the uniform vector potential for a monochromatic laser, and
$\bm{R}_{ij}=\bm{R}_{i}-\bm{R}_{j}$
with $\bm{R}_{i}$ being the location of the $i$th site.
With the Jacobi-Anger identity, the hopping amplitude is expanded in a Fourier series as
\begin{equation}
t_{ij}(t)=\sum_{m=-\infty}^{\infty}t_{ij}^{(m)}e^{-im\omega t}
=\sum_{m=-\infty}^{\infty}\left[t_{ij}\mathcal{J}_{m}(\alpha_{ij})(-i)^{m}e^{im\theta_{ij}}\right]e^{-im\omega t},
\end{equation}
where $\mathcal{J}_m$ is the $m$th Bessel function, and we have denoted $\bm{A}\cdot\bm{R}_{ij}\equiv \alpha_{ij}e^{i\theta_{ij}}$.

For later convenience we decompose the Hamiltonian as $\hat{H}_\text{Hub}(t)=-\lambda\hat{T}(t)+U\hat{D}$,
where $\hat{T}(t)$ is the kinetic energy operator, 
$\hat{D}=(1/2)\sum_{i}\hat{n}_{i}(\hat{n}_{i}-1)$ is the double occupancy operator, and
$\lambda=1$ is a bookkeeping parameter for the perturbative expansion,
i.e., we formally consider an expansion in $\lambda$.

\subsection{Strong-coupling expansion}

In order to derive the equation of motion for the spin degree of freedom, 
here we introduce a time-dependent canonical transformation $e^{i\hat{\mathcal{S}}(t)}$ 
to perform the strong-coupling expansion~\cite{Bukov2016,Kitamura2016,Claassen2016} first, 
prior to the high-frequency expansion as we stressed in the Introduction. 
The time-dependent Schr\"odinger equation, 
$i\partial_{t}|\Psi(t)\rangle=\hat{H}_{\text{Hub}}(t)|\Psi(t)\rangle$, 
can then be expressed for the transformed Hamiltonian as
\begin{equation}
\hat{H}_{\text{SCE}}(t)+U\hat{D}
=e^{i\hat{\mathcal{S}}(t)}[\hat{H}_{\text{Hub}}(t)-i\partial_{t}]e^{-i\hat{\mathcal{S}}(t)}
\label{eq:sce-def}
\end{equation}
for the transformed state vector $|\Phi(t)\rangle=e^{i\hat{\mathcal{S}}(t)}|\Psi(t)\rangle$.
In this definition we have separated $U\hat{D}$ 
from the strong-coupling Hamiltonian $\hat{H}_{\text{SCE}}(t)$
to simplify Eq.~(\ref{eq:sce-recursive}) below.

Now we aim to express $|\Phi(t)\rangle$ in terms of spin configurations 
that span the $\hat{D}=0$ subspace.
This can be done if $\hat{H}_{\text{SCE}}(t)$ is diagonal in $\hat{D}$
(i.e., $\hat{\mathcal{S}}(t)$ block-diagonalizing the Hamiltonian), 
and if $U$ is so large that $\hat{D}\neq0$ sectors can be neglected for a low-energy description. 
Namely, to obtain the spin Hamiltonian, we should determine $\hat{\mathcal{S}}(t)$
to eliminate the terms offdiagonal in $\hat{D}$ by expanding in $\lambda$ as 
$\hat{H}_\text{SCE}(t)=\sum_{n=1}^\infty\lambda^n\hat{H}_\text{SCE}^{(n)}$ and 
$\hat{\mathcal{S}}(t)=\sum_{n=1}^\infty\lambda^n\hat{\mathcal{S}}^{(n)}$.
For this we can rewrite Eq.~(\ref{eq:sce-def}) as
\begin{multline}
\hat{H}_{\text{SCE}}(t)+U[\hat{D},i\hat{\mathcal{S}}(t)]+\partial_{t}\hat{\mathcal{S}}(t)\\
=-\lambda\hat{T}(t)-\sum_{n=1}^{\infty}\dfrac{B_{n}}{n!}\text{ad}_{i\hat{\mathcal{S}}}^{n}
\left((-1)^{n}\lambda\hat{T}(t)+\hat{H}_{\text{SCE}}(t)\right),
\label{eq:sce-recursive}
\end{multline}
where $B_n$ is the Bernoulli number and $\text{ad}_XY=[X,Y]$.

Since $\hat{H}_\text{SCE}(t)=\mathcal{O}(\lambda)$ and
$\hat{\mathcal{S}}(t)=\mathcal{O}(\lambda)$,
the $N$th order term on the right-hand side is composed of
$\hat{H}_\text{SCE}^{(n)}$ and  $\hat{\mathcal{S}}^{(n)}$ with $n\le N-1$,
while the left-hand side with $n=N$. 
Namely, we can determine the form of $\hat{H}_\text{SCE}$ and $\hat{\mathcal{S}}$
order by order with this expression.
If we denote
$\hat{R}^{(n)}\equiv \hat{H}_{\text{SCE}}^{(n)}+[U\hat{D},i\hat{\mathcal{S}}^{(n)}]
+\partial_{t}\hat{\mathcal{S}}^{(n)}$,
they are expressed up to the fourth order as
\begin{align}
\hat{R}^{(1)} & =-\hat{T},\label{eq:recursive-1}\\
\hat{R}^{(2)} & =-\dfrac{1}{2}[i\hat{\mathcal{S}}^{(1)},\hat{T}-\hat{H}_{\text{SCE}}^{(1)}],
\label{eq:recursive-2}\\
\hat{R}^{(3)} & =-\dfrac{1}{2}[i\hat{\mathcal{S}}^{(2)},\hat{T}-\hat{H}_{\text{SCE}}^{(1)}]
+\dfrac{1}{2}[i\hat{\mathcal{S}}^{(1)},\hat{H}_{\text{SCE}}^{(2)}]
\nonumber\\
 & -\dfrac{1}{12}[i\hat{\mathcal{S}}^{(1)},[i\hat{\mathcal{S}}^{(1)},\hat{T}+\hat{H}_{\text{SCE}}^{(1)}]],
 \label{eq:recursive-3}\\
\hat{R}^{(4)} & =-\dfrac{1}{2}[i\hat{\mathcal{S}}^{(3)},\hat{T}-\hat{H}_{\text{SCE}}^{(1)}]
+\dfrac{1}{2}[i\hat{\mathcal{S}}^{(2)},\hat{H}_{\text{SCE}}^{(2)}]
+\dfrac{1}{2}[i\hat{\mathcal{S}}^{(1)},\hat{H}_{\text{SCE}}^{(3)}]
\nonumber \\
 & -\dfrac{1}{12}[i\hat{\mathcal{S}}^{(1)},[i\hat{\mathcal{S}}^{(1)},\hat{H}_{\text{SCE}}^{(2)}]]
 -\dfrac{1}{12}[i\hat{\mathcal{S}}^{(1)},[i\hat{\mathcal{S}}^{(2)},\hat{T}+\hat{H}_{\text{SCE}}^{(1)}]]
\nonumber \\
 & -\dfrac{1}{12}[i\hat{\mathcal{S}}^{(2)},[i\hat{\mathcal{S}}^{(1)},\hat{T}+\hat{H}_{\text{SCE}}^{(1)}]]
\label{eq:recursive-4}.
\end{align}

The operator $\hat{\mathcal{S}}$ should be determined such that
$\hat{H}_\text{SCE}$ be block-diagonal.
While in general $\hat{H}_\text{SCE}$ is not necessarily time-periodic~\footnote{
For example, one can choose a boundary condition for the time axis in such a way that 
the transformation coincides with that in the equilibrium case 
at the time at which the electric field is turned on.},
it is, as seen from Eq.~(\ref{eq:sce-def}), if the transformation $\hat{\mathcal{S}}(t)$ is time-periodic.
Such a time-periodic transformation can be uniquely determined
if one imposes that $\hat{\mathcal{S}}(t)$ does not contain block-diagonal terms.
In such a case, one can decompose the transformation into Fourier components of block matrices as
\begin{equation}
\hat{\mathcal{S}}^{(n)}=\sum_{d\neq0}\sum_{m=-\infty}^{\infty}\hat{\mathcal{S}}_{d,m}^{(n)}e^{-im\omega t},
\label{eq:is-decomp}
\end{equation}
where ${\mathcal{S}}_{d,m}^{(n)}$ changes the double occupancy by $d$; 
$[\hat{D},\hat{\mathcal{S}}_{d,m}^{(n)}]=d\hat{\mathcal{S}}_{d,m}^{(n)}$.
By decomposing Eq.~(\ref{eq:sce-recursive}) in terms of $d$ and $m$
as well and dividing both sides by $(dU-m\omega)$,
one obtains explicit expressions for ${\mathcal{S}}_{d,m}^{(n)}$ 
and $\hat{H}_\text{SCE}^{(n)}$, as we shall see below.

\subsection{Second-order perturbation\label{sec:2nd}}

We can now give the explicit form of the effective spin Hamiltonian.
Let us first decompose the hopping operator as
\begin{equation}
\hat{T}(t)=\sum_{m=-\infty}^{\infty}(\hat{T}_{-1,m}
+\hat{T}_{0,m}+\hat{T}_{+1,m})e^{-im\omega t},
\end{equation}
where $\hat{T}_{d,m}$ changes the double occupancy by $d$.
As we are interested in spin correlations up to three bodies,
we neglect terms involving more than four sites.
In such a situation, eigenvalues of the double occupancy operator $\hat{D}$ takes 0 or 1,
so that one can restrict the range of the $d$ summation in Eq.~(\ref{eq:is-decomp}) to $d=\pm1$
\footnote{
We can note that the present formalism is so general that it is applicable 
to SU($N$)-Hubbard model straightforwardly, 
in which case additional terms $\hat{T}_{\pm2,m}$ and $\hat{\mathcal{S}}_{\pm2,m}^{(n)}$ appear.
}.

Then Eqs.~(\ref{eq:recursive-1}) and (\ref{eq:recursive-2}) give
\begin{align}
\hat{H}_{\text{SCE}}^{(1)} & =-\sum_{m=-\infty}^{\infty}\hat{T}_{0,m}e^{-im\omega t},\\
i\hat{S}^{(1)} & =-\sum_{m=-\infty}^{\infty}\dfrac{\hat{T}_{+1,m}}{(U-m\omega)}e^{-im\omega t}-\text{H.c.},\\
\hat{H}_{\text{SCE}}^{(2)} & =\sum_{n,m=-\infty}^{\infty}\dfrac{[\hat{T}_{+1,n},\hat{T}_{-1,m-n}]}{2(U-n\omega)}e^{-im\omega t}+\text{H.c.},\\
i\hat{S}^{(2)} & =\sum_{n,m=-\infty}^{\infty}\dfrac{[\hat{T}_{+1,n},\hat{T}_{0,m-n}]}{(U-n\omega)(U-m\omega)}e^{-im\omega t}-\text{H.c.}
\end{align}

The expanded Hamiltonian is block-diagonal in $\hat{D}$
up to the truncation order, so that the $\hat{D}=0$ block,
$\hat{P}_0\hat{H}_\text{SCE}\hat{P}_0$
with $\hat{P}_0$ being the projection onto the $\hat{D}=0$ sector,
gives the spin Hamiltonian.
The explicit form in terms of the spin operator is efficiently obtained by the following procedure:

(1) We first expand the commutators and express
$\hat{P}_0\hat{H}_\text{SCE}\hat{P}_0$ in terms of products of $\hat{T}_{d,m}$ 
as 
$\hat{P}_0\hat{H}_{\text{SCE}}^{(1)}\hat{P}_0 =0$,
$\hat{P}_0\hat{H}_{\text{SCE}}^{(2)}\hat{P}_0 =
-\sum_{n,m}\hat{T}_{-1,m-n}\hat{T}_{+1,n}e^{-im\omega t}/(2(U-n\omega))+\text{H.c.}$

\begin{figure}
\begin{centering}
\includegraphics{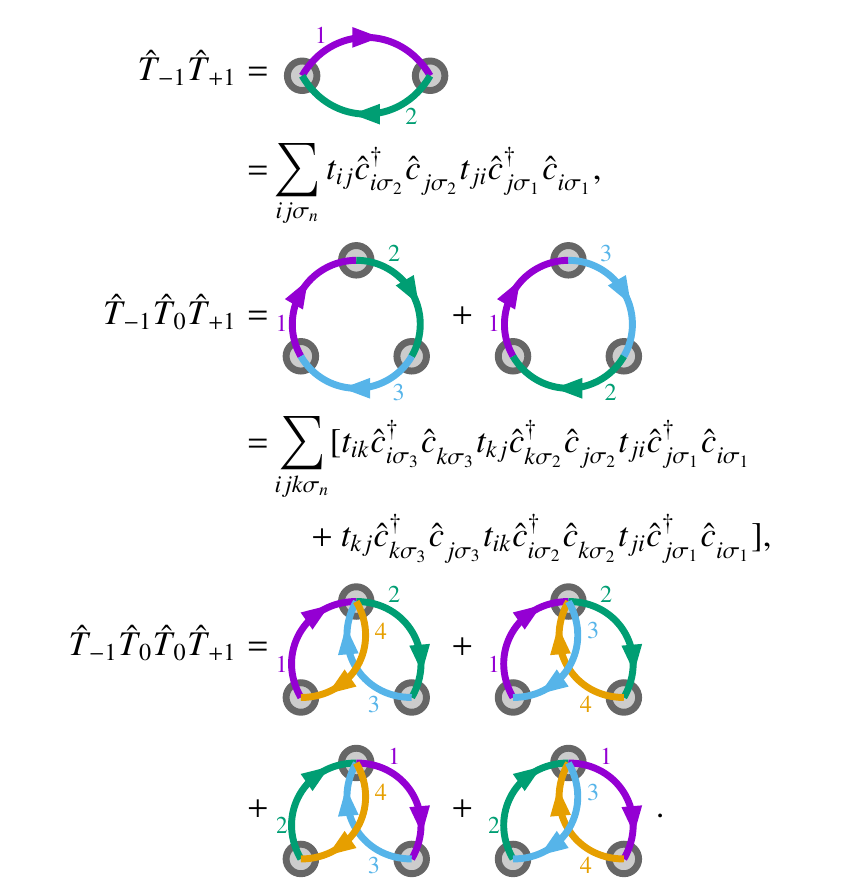}
\caption{Diagrammatic representation of products of hopping operators.
Gray circles represent sites,
while arrows represent hopping of electrons. Numbers represent the order of the hopping processes.
\label{fig:sce-diagram}}
\end{centering}
\end{figure}

(2) We then represent the operator products diagrammatically as in Fig.~\ref{fig:sce-diagram}, 
and we express them in terms of the original electron 
creation and annihilation operators as 
$\hat{T}_{-1,m-n}\hat{T}_{+1,n}
= \sum_{ij}\sum_{\sigma_1\sigma_2}t_{ji}^{(m-n)}\hat{c}^\dagger_{i\sigma_2}\hat{c}_{j\sigma_2}
t_{ij}^{(n)}\hat{c}^\dagger_{j\sigma_1}\hat{c}_{i\sigma_1}$.

(3) If we rearrange the electron operators, 
we can express these terms with cyclic permutation operators,
\begin{align}
\hat{\mathcal{P}}_{ij} & =\sum_{\sigma_{1},\sigma_{2}}
\hat{c}_{i\sigma_{2}}^{\dagger}\hat{c}_{i\sigma_{1}}^{\phantom{\dagger}}
\hat{c}_{j\sigma_{1}}^{\dagger}\hat{c}_{j\sigma_{2}}^{\phantom{\dagger}},\\
\hat{\mathcal{P}}_{ijk} & =\sum_{\sigma_{1},\sigma_{2},\sigma_{3}}
\hat{c}_{i\sigma_{3}}^{\dagger}\hat{c}_{i\sigma_{1}}^{\phantom{\dagger}}
\hat{c}_{j\sigma_{1}}^{\dagger}\hat{c}_{j\sigma_{2}}^{\phantom{\dagger}}
\hat{c}_{k\sigma_{2}}^{\dagger}\hat{c}_{k\sigma_{3}}^{\phantom{\dagger}}.
\end{align}
Namely, $\hat{T}_{-1,m-n}\hat{T}_{+1,n}
= \sum_{ij}t_{ji}^{(m-n)}t_{ij}^{(n)}(1-\hat{\mathcal{P}}_{ij})$, etc.

(4) Finally we express the permutation operators in terms of the spin operators;
\begin{gather}
\hat{\mathcal{P}}_{ij}=2\hat{\bm{S}}_{i}\cdot\hat{\bm{S}}_{j}+\frac{1}{2},\\
\hat{\mathcal{P}}_{ijk}-\hat{\mathcal{P}}_{ijk}^{\dagger}
=-4i(\hat{\bm{S}}_{i}\times\hat{\bm{S}}_{j})\cdot\hat{\bm{S}}_{k},\\
\hat{\mathcal{P}}_{ijk}+\hat{\mathcal{P}}_{ijk}^{\dagger}
=2(\hat{\bm{S}}_{i}\cdot\hat{\bm{S}}_{j}+\hat{\bm{S}}_{j}\cdot
\hat{\bm{S}}_{k}+\hat{\bm{S}}_{k}\cdot\hat{\bm{S}}_{i})+\dfrac{1}{2},
\end{gather}
where
\begin{equation}
\hat{\bm{S}}_{i}=\dfrac{1}{2}\sum_{\sigma_{1}\sigma_{2}}
\hat{c}_{i\sigma_{1}}^{\dagger}\bm{\sigma}_{\sigma_{1}\sigma_{2}}^{\phantom{\dagger}}
\hat{c}_{i\sigma_{2}}^{\phantom{\dagger}},
\end{equation}
with $\bm{\sigma}$ being the 1/2-spin Pauli matrix on $i$th
site.

With this prescription, the leading-order spin Hamiltonian reduces to
\begin{align}
\hat{H}_{\text{spin}}(t)=\hat{P}_0\hat{H}_{\text{SCE}}(t)\hat{P}_0 &
=\lambda^2\sum_{\langle i,j\rangle}J_{ij}(t)
\left(\hat{\bm{S}}_{i}\cdot\hat{\bm{S}}_{j}-\frac{1}{4}\right)
\label{eq:td-heisenberg},
\end{align}
where the time-periodic exchange interaction is expressed as
\begin{equation}
J_{ij}(t)  =\sum_{n,m=-\infty}^{\infty}(-1)^{m}
\frac{4|t_{ij}|^2\mathcal{J}_{n+m}(\alpha_{ij})\mathcal{J}_{n-m}(\alpha_{ij})}{U-(n+m)\omega}
\cos{2m(\omega t-\theta_{ij})}.
\label{eq:td-exchange}
\end{equation}

\subsection{High-frequency expansion}
Before going over to higher-order terms in the strong-coupling expansion,
let us first look at how the time-dependent Heisenberg model, 
Eq.~(\ref{eq:td-heisenberg}), behaves when the driving frequency is larger than
the Fourier components of the exchange intercation, Eq.~(\ref{eq:td-exchange}).
In this situation, 
one can perform a high-frequency expansion~\cite{Casas2001,Bukov2015,Eckardt2015}
to derive the effective static Hamiltonian.
For this, we introduce another time-periodic canonical transformation $\hat{\Lambda}(t)$ to obtain
\begin{equation}
\hat{F}_\text{spin}
=e^{i\hat{\Lambda}(t)}(\hat{H}_\text{spin}(t)-i\partial_t)e^{-i\hat{\Lambda}(t)},
\end{equation}
to eliminate the time dependence of the Hamiltonian 
in an expansion in $1/\omega$.
If we denote the $m$th Fourier component of $\hat{H}_\text{spin}(t)$
as $\hat{H}_{\text{spin},m}$,
the effective static Hamiltonian (quasienergy operator) $\hat{F}_\text{spin}$
is obtained as~\cite{Casas2001,Bukov2015,Eckardt2015}
\begin{equation}
\hat{F}_\text{spin}=\hat{H}_{\text{spin},0}
-\sum_{m\neq0}\frac{[\hat{H}_{\text{spin},m},\hat{H}_{\text{spin},-m}]}{2m\omega}
+\mathcal{O}(\omega^{-2}).\label{eq:hfe}
\end{equation}

With an identity
\begin{equation}
[\hat{\mathcal{P}}_{ij},\hat{\mathcal{P}}_{jk}]
=\hat{\mathcal{P}}_{ijk}-\hat{\mathcal{P}}_{ijk}^{\dagger},
\end{equation}
we obtain the effective static Hamiltonian as a Heisenberg model
that has an emergent scalar chirality term as
\begin{equation}
\hat{F}_\text{spin}=\lambda^2\sum_{\langle i,j\rangle}
J_{ij}^{\text{(h)}}\left(\hat{\bm{S}}_{i}\cdot\hat{\bm{S}}_{j}
-\frac{1}{4}\right)+\lambda^4\sum_{ijk}
J_{\chi,ijk}^{\text{(h)}}(\hat{\bm{S}}_{i}\times\hat{\bm{S}}_{j})\cdot\hat{\bm{S}}_{k},
\end{equation}
where
\begin{align}
J_{ij}^{\text{(h)}} =&\sum_{n=-\infty}^{\infty}
\frac{4|t_{ij}|^2\mathcal{J}_{n}^2(\alpha_{ij})}{U-n\omega},
\label{eq:eff-heisenberg}
\end{align}
and the  scalar chirality coefficient 
\begin{align}
J_{\chi,ijk}^{\text{(h)}}=&
-\sum_{m=1}^{\infty}8|t_{ij}|^{2}|t_{jk}|^{2}U^{2}\sin2m(\theta_{ij}-\theta_{jk})
\nonumber\\
\times &
\sum_{n,l=-\infty}^{\infty}
\frac{\mathcal{J}_{n+m}(\alpha_{ij})\mathcal{J}_{n-m}(\alpha_{ij})
\mathcal{J}_{l+m}(\alpha_{jk})\mathcal{J}_{l-m}(\alpha_{jk})}
{m\omega[U^{2}-(n+m)^{2}\omega^{2}][U^{2}-(l+m)^{2}\omega^{2}]}
\label{eq:eff-chiral-1}.
\end{align}
Note that, while the exchange coupling $J_{ij}^\text{(h)}$ is defined 
for bonds $\langle i,j\rangle$
(which coincides with that obtained originally in Ref.~\cite{Mentink2014}), 
the chiral coupling
$J_{\chi,ijk}^{\text{(h)}}$ is defined here for each sequence $ijk$,
i.e., the coupling constant for a given combination of three sites $\{i,j,k\}$ is given as
a sum of six different permutations, 
$J_{\chi,ijk}^\text{(h)}+J_{\chi,kij}^\text{(h)}+\cdots$.

These expressions should be accurate only 
when the driving frequency $\omega$ is much smaller than $U$.
Otherwise, the higher-order terms in the strong-coupling expansion should become relevant,
since they also include a fourth-order contribution in $\lambda$.
We note that the high-frequency expansion introduced here,
justified for $\omega\ll U$,
is quite \textit{opposite} to that of the periodically driven Hubbard model 
described by Eq.~(\ref{eq:td-hubbard}),
which requires $U\ll\omega$.
Here ``high-frequency" means that the frequency is larger than $J_{ij}$,
since the expansion is in the Fourier components of $J_{ij}$ devided by $\omega$. 

In the leading order in the driving amplitude, 
the chiral coupling $J_{\chi,ijk}^{\text{(h)}}$ behaves as
\begin{equation}
J_{\chi,ijk}^{\text{(h)}}\sim
-\frac{|t_{ij}|^{2}|t_{jk}|^{2}
(U^{2}+2\omega^{2})^{2}\omega^{3}\alpha_{ij}^{2}\alpha_{jk}^{2}}
{2 U^{2}(U^{2}-\omega^{2})^{2}(U^{2}-4\omega^{2})^{2}}
\sin2(\theta_{ij}-\theta_{jk}),
\end{equation}
which is quartic in the electric field strength,
so that this is certainly a nonlinear response.
By contrast, 
the emergent scalar chirality is obtained as a linear response to the laser's field 
due to the higher-order terms in the strong-coupling expansion, as we shall see below.

\subsection{Fourth-order perturbation}
Let us continue our formalism for the strong-coupling expansion to fourth order,
which has to be complemented by the terms not contained in the above high-frequency expansion 
to complete the expression for chiral coupling 
with respect to the leading order in $\lambda$. 
The third-order perturbation Eq.~(\ref{eq:recursive-3}) yields
\begin{align}
\hat{H}_{\text{SCE}}^{(3)}=-\sum_{l,n,m=-\infty}^{\infty}&
\frac{[[\hat{T}_{+1,l},\hat{T}_{0,n-l}],\hat{T}_{-1,m-n}]}
{2(U-l\omega)(U-n\omega)}e^{-im\omega t}
+\text{H.c.},
\end{align}
\begin{align}
i\hat{\mathcal{S}}^{(3)}=-\sum_{l,n,m=-\infty}^{\infty}\biggl[&
\frac{[\hat{T}_{+1,l},[\hat{T}_{+1,n},\hat{T}_{-1,m-n-l}]]}
{3(U-l\omega)(U-n\omega)(U-m\omega)}
\nonumber\\
+&\frac{[\hat{T}_{-1,l},[\hat{T}_{+1,n},\hat{T}_{-1,m-n-l}]]}
{3(U+l\omega)(U-n\omega)(U+m\omega)}
\nonumber\\
+&\frac{[[\hat{T}_{+1,l},\hat{T}_{0,n-l}],\hat{T}_{0,m-n}]}
{(U-l\omega)(U-n\omega)(U-m\omega)}
\biggr]e^{-im\omega t}
-\text{H.c.}
\end{align}

By following the prescription given in Sec.~\ref{sec:2nd},
we obtain the third-order terms for the spin Hamiltonian as
\begin{multline}
\hat{H}_{\text{spin}}^{(3)}=\sum_{ijk}\Lambda_{ijk}(t)\biggl[
\hat{\bm{S}}_{i}\cdot\hat{\bm{S}}_{j}+\hat{\bm{S}}_{j}\cdot\hat{\bm{S}}_{k}
-\hat{\bm{S}}_{k}\cdot\hat{\bm{S}}_{i}\\
-2i(\hat{\bm{S}}_{i}\times\hat{\bm{S}}_{j})\cdot\hat{\bm{S}}_{k}
-\dfrac{1}{4}\biggr]+\text{H.c.},
\label{H3spin}
\end{multline}
where
\begin{multline}
\Lambda_{ijk}(t)=\sum_{l,n,m=-\infty}^{\infty}
\frac{i\mathcal{J}_{l}(\alpha_{ij})\mathcal{J}_{m-n}(\alpha_{jk})\mathcal{J}_{n-l}(\alpha_{ki})}
{(U-l\omega)(U-n\omega)}\\
\times \text{Im}[t_{ij}t_{jk}t_{ki}(-i)^{m}]
e^{-im(\omega t-\theta_{jk})-in(\theta_{jk}-\theta_{ki})-il(\theta_{ki}-\theta_{ij})}.
\end{multline}
While Eq.~(\ref{H3spin}) contains a scalar chirality term,
it has no static component ($m=0$) as seen from the expression 
unless the bare hopping amplitude is complex.
Oscillating components ($m\neq0$) give an $\mathcal{O}(\lambda^5)$ contribution
in the high-frequency expansion.

As for the chiral coupling $J_{\chi,ijk}$, 
higher-order contributions emerge as the time average of the fourth-order perturbation, Eq.~(\ref{eq:recursive-4}), and we have
\begin{align}
\overline{\hat{H}_{\text{SCE}}^{(4)}(t)}=\sum_{l,n,m=-\infty}^{\infty}\biggl[&
\dfrac{[\hat{T}_{-1,-m},[\hat{T}_{+1,l},[\hat{T}_{-1,n},\hat{T}_{+1,m-n-l}]]]}
{8(U-l\omega)(U+n\omega)(U-m\omega)}
\nonumber\\
-&\dfrac{[\hat{T}_{-1,-m},[\hat{T}_{+1,l},[\hat{T}_{+1,n},\hat{T}_{-1,m-n-l}]]]}
{8(U-l\omega)(U-n\omega)(U-m\omega)}
\nonumber\\
+&\dfrac{[[[\hat{T}_{+1,l},\hat{T}_{0,n-l}],\hat{T}_{0,m-n}],\hat{T}_{-1,-m}]}
{2(U-l\omega)(U-n\omega)(U-m\omega)}\biggr]+\text{H.c.}
\end{align}
With this term, we arrive at an effective static Hamiltonian
$\hat{F}_\text{spin}$ as
\begin{equation}
\hat{F}_\text{spin}=\lambda^2\sum_{\langle i,j\rangle}
J_{ij}^{\text{(h)}}\left(\hat{\bm{S}}_{i}\cdot\hat{\bm{S}}_{j}
-\frac{1}{4}\right)+\lambda^4\sum_{ijk}
J_{\chi,ijk}(\hat{\bm{S}}_{i}\times\hat{\bm{S}}_{j})\cdot\hat{\bm{S}}_{k},
\end{equation}
with
\begin{equation}
J_{\chi,ijk}=J_{\chi,ijk}^\text{(h)}+J_{\chi,ijk}^\text{(c)},
\end{equation}
\begin{widetext}
\begin{multline}
J_{\chi,ijk}^{\text{(c)}}=-4|t_{ij}|^{2}|t_{jk}|^{2}\sum_{l,n,m=-\infty}^{\infty}\biggl[
\frac{\mathcal{J}_{l}(\alpha_{ij})\mathcal{J}_{n}(\alpha_{jk})
\mathcal{J}_{l+m}(\alpha_{ij})\mathcal{J}_{n+m}(\alpha_{jk})
\sin m(\theta_{ij}-\theta_{jk})}
{(U-l\omega)(U-n\omega)(U-(l+n+m)\omega)}\\
+\dfrac{\mathcal{J}_{l+m}(\alpha_{ij})\mathcal{J}_{l-m}(\alpha_{ij})
\mathcal{J}_{n+m}(\alpha_{jk})\mathcal{J}_{n-m}(\alpha_{jk})
\sin2m(\theta_{ij}-\theta_{jk})}
{(U-(l-m)\omega)(U-(l+m)\omega)(U-(n+m)\omega)}\biggr]
\label{eq:eff-chiral-2}.
\end{multline}
\end{widetext}
The same expression was also obtained in a recent study~\cite{Claassen2016},
where the effective Hamiltonian was derived 
by simultaneously eliminating the offdiagonal terms in $\hat{D}$ and the time dependence.
The present formalism considers the strong-coupling and high-frequency expansion separately 
to perform them in turn.
This clarifies that the high-frequency expansion is applicable 
when the driving frequency is larger than 
the effective exchange interaction rather than the hopping or 
on-site interaction, if the first 
strong-coupling expansion Eq.~(\ref{eq:sce-recursive}) is justified.
The strong-coupling expansion is justified when the denominator $(U-n\omega)$ is sufficiently large
or the numerator $\propto\mathcal{J}_n(A)$ is small.
The present formalism can be utilized for the low-frequency ($\omega<J$) driving, if one only performs 
the strong-coupling expansion to analyze the time-periodic spin Hamiltonian $H_\text{spin}(t)$.

Further, we can notice that the first term in Eq.~(\ref{eq:eff-chiral-2}) 
has a contribution \textit{quadratic} (as opposed to quartic) in the field amplitude, 
namely
\begin{equation}
J_{\chi,ijk}^{\text{(c)}}\sim\dfrac{|t_{ij}|^{2}|t_{jk}|^{2}\omega(7U^{2}-3\omega^{2})}
{U^{2}(U^2-\omega^2)^{3}}
i(\bm{E}^{\ast}\times\bm{E})\cdot(\bm{R}_{ij}\times\bm{R}_{jk}),
\end{equation}
where $\bm{E}=i\omega\bm{A}$.
This implies that the scalar spin chirality affects the optical responses of Mott insulators
even in the linear-response regime. We will elaborate upon this observation later in the paper.

\section{Comparison with numerical calculations for circularly polarized lasers}
Here we verify the validity of the expansion 
by comparing the results obtained from the effective Hamiltonian
with numerically exact spectra of a periodically driven Hubbard cluster.  
Specifically, a circularly polarized laser with a 
vector potential $\bm{A}=A(1,i)$ 
is considered for a three-site Hubbard model in an equilateral triangular geometry 
with hopping amplitudes being unity. 
We note that the numerical calculation should contain finite-size effects, 
so we have to be careful in comparing the results with the perturbative expansion. 
However, when the expansion is justified (i.e., when the system is a Mott insulator), 
electrons cannot hop to distant sites in virtual processes,
so that the finite-size effect should be small.
In the following, we first discuss how to extract the coupling constants from numerical calculations,
and then we compare them with those obtained by the perturbative expansion.

\subsection{Numerical computation}
A direct way to numerically calculate the effective spin Hamiltonian could be 
a continuous unitary transformation (also known as Wegner's flow-equation approach) 
extended to periodically driven systems~\cite{KehreinBook,Verdeny2013}.
Namely, one parametrizes the unitary transformation $e^{i\hat{\mathcal{S}}(t)}$
by the flow parameter $l$ to define the transformation with a differential equation,
\begin{equation}
\partial_{l}\hat{U}(l,t)=\hat{\eta}(l,t)\hat{U}(l,t).
\end{equation}
The generator of the transformation, $\hat{\eta}(l,t)$, can here be chosen as
$\hat{\eta}(l,t)=[U\hat{D},\hat{F}(l,t)]$ for the strong-coupling expansion, 
or $\hat{\eta}(l,t)=-i\partial_{t}\hat{F}(l,t)$ for the high-frequency expansion. 
Taking  $l\rightarrow\infty$ yields the desired effective Hamiltonian.

Here, let us employ an alternative way with smaller numerical costs,
i.e., the numerical construction of the Floquet-Magnus Hamiltonian. 
As we have seen, if $|\Psi(t)\rangle$ satisfies the time-dependent Schr\"odinger equation,
$e^{i\hat{\Lambda}(t)}\hat{P}_0e^{i\hat{\mathcal{S}}(t)}|\Psi(t)\rangle$ satisfies a 
static Schr\"odinger equation with $\hat{F}_\text{spin}$.
Namely, the time-evolution operator $\hat{\mathcal{U}}(t,t_0)$ can be represented as
\begin{equation}
\hat{\mathcal{U}}(t,t_0) =
e^{-i\hat{\mathcal{S}}(t)}\hat{P}_0e^{-i\hat{\Lambda}(t)}e^{-i\hat{F}_\text{spin}(t-t_0)}
e^{i\hat{\Lambda}(t_0)}\hat{P}_0e^{i\hat{\mathcal{S}}(t_0)}.
\end{equation}
Since $\hat{\mathcal{S}}(t)$ and $\hat{\Lambda}(t)$ are time-periodic,
one can express the effective Hamiltonian as
\begin{equation}
\hat{F}_\text{spin}=\frac{i\omega}{2\pi}
e^{i\hat{\Lambda}(t_0)}\hat{P}_0e^{i\hat{\mathcal{S}}(t_0)}
\ln\hat{\mathcal{U}}\left(t_0+\frac{2\pi}{\omega},t_0\right)
e^{-i\hat{\mathcal{S}}(t_0)}\hat{P}_0e^{-i\hat{\Lambda}(t_0)}
\end{equation}
up to modulo $\omega$ (due to the indefiniteness of the logarithm).

While the construction of the transformation
$e^{i\hat{\Lambda}(t_0)}\hat{P}_0e^{i\hat{\mathcal{S}}(t_0)}$ is numerically demanding,
the time-evolution operator can be calculated more easily
by numerically integrating the equation of motion
$i\partial_t\hat{\mathcal{U}}\left(t,t_0\right)=\hat{H}_\text{Hub}(t)\hat{\mathcal{U}}(t,t_0)$.
Hence diagonalization of $(i\omega/2\pi)\ln\hat{\mathcal{U}}(t_0+2\pi/\omega,t_0)$
yields the eigenvalues (quasienergy, energy up to modulo $\omega$) of $\hat{F}_\text{spin}$
(along with those of $\hat{D}\neq0$ sectors,
which can be distinguished
by checking the time average of the expectation value of the double occupancy).

Since the three-site spin model, 
$\hat{F}_\text{spin}=J(\hat{\bm{S}}_{1}\cdot\hat{\bm{S}}_{2}
+\hat{\bm{S}}_{2}\cdot\hat{\bm{S}}_{3}
+\hat{\bm{S}}_{3}\cdot\hat{\bm{S}}_{1}-3/4)
+J_{\chi}(\hat{\bm{S}}_{1}\times\hat{\bm{S}}_{2})\cdot\hat{\bm{S}}_{3}$, 
has quadruply degenerated $0$ and
doubly degenerated $-6J\pm2\sqrt{3}J_\chi$ as eigenvalues,
one can reconstruct $J$ and $J_\chi$ from the computed eigenvalues
without calculating the transformation
$e^{i\hat{\Lambda}(t_0)}\hat{P}_0e^{i\hat{\mathcal{S}}(t_0)}$.

\subsection{Behavior of the coupling constants}
\begin{figure}
\begin{centering}
\includegraphics{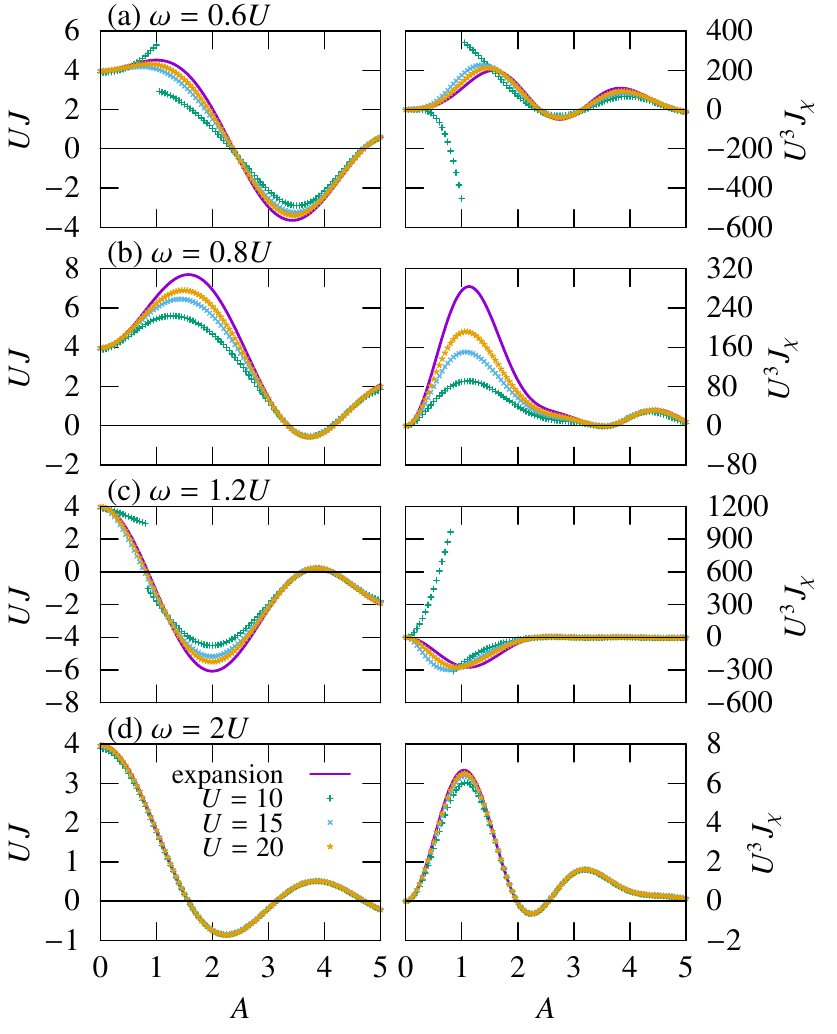}
\caption{Left: The exchange interaction $J$ 
normalized by $1/U$ against the intensity, $A$, of the periodic drive for four cases of $\omega/U=0.6$-$2.0$. 
Right: the chiral coupling $J_{\chi}$ normalized by $1/U^3$. Symbols with different colors represent the numerical results for the three-site driven Hubbard model for several values of $U=10-20$,
while solid curves (purple) the perturbative expression
obtained by the strong-coupling expansion,
\label{fig:repulsive-params}}
\end{centering}
\end{figure}

We can thus compare the behavior of the coupling constants.
As Eqs.~(\ref{eq:eff-heisenberg}),~(\ref{eq:eff-chiral-1}), and (\ref{eq:eff-chiral-2}) indicate,
$UJ$ and $U^3J_\chi$ should be functions of $\omega/U$ in the leading-order,
so that here we look at 
how the numerical result for these rescaled couplings changes with $U$ for a fixed $\omega/U$.
Figure~\ref{fig:repulsive-params} displays $UJ$ and $U^3J_\chi$ against the intensity, $A$, of the periodic drive 
for various values of $\omega/U$.
We can see that the perturbative expression agrees well with 
the numerical results, where the agreement becomes 
improved for increasing $U$ in all the cases.

\begin{figure}
\begin{centering}
\includegraphics{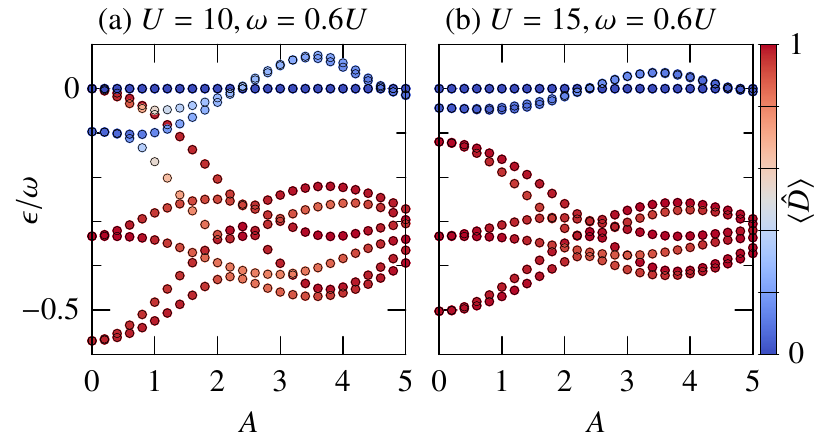}
\caption{Quasienergy spectra, normalized by $\omega$, 
are plotted against the driving amplitude $A$ for the three-site Hubbard model 
driven by a circularly polarized laser for 
(a) $U=10$, $\omega=0.6U=6$ or 
(b) $U=15$, $\omega=0.6U=9$.
Color code represents the time-averaged expectation value
of the double occupancy $\hat{D}$ for each quasienergy eigenstate.
\label{fig:docc-mixing}}
\end{centering}
\end{figure}

However, we also notice that coupling constants can become discontinuous
against $A$ when $U$ is small.
In particular, the chiral coupling $J_\chi$ deviates significantly 
from the perturbative result even qualitatively,
which signals a breakdown of the strong-coupling expansion.
We can indeed trace back its origin to a level crossing
between $\hat{D}=0$ and $\hat{D}=1$ sectors of $\hat{F}_\text{spin}$,
for which the strong-coupling expansion becomes inapplicable. 
Let us show an example of the quasienergy spectra 
along with the time-averaged expectation value of the double occupancy
in Fig.~\ref{fig:docc-mixing} for $\omega=0.6U$.  
For a relatively large $U=15$, quasienergy eigenstates can be 
unambiguously classified into $\langle\hat{D}\rangle\sim0$ and $1$, 
which can be transformed into $\hat{D}=0$ and $1$ sectors
in the perturbative transformation.
For a smaller $U=10$, on the other hand, 
level crossings between $\hat{D}=0$, $1$ sectors set in around $A\sim1$.
There, $\langle\hat{D}\rangle\sim0$ states (doubly degenerate) continuously change into
those with $\langle\hat{D}\rangle\sim1$ when $A$ is increased, 
which cannot be described by the perturbative transformation.

\begin{figure}
\begin{centering}
\includegraphics{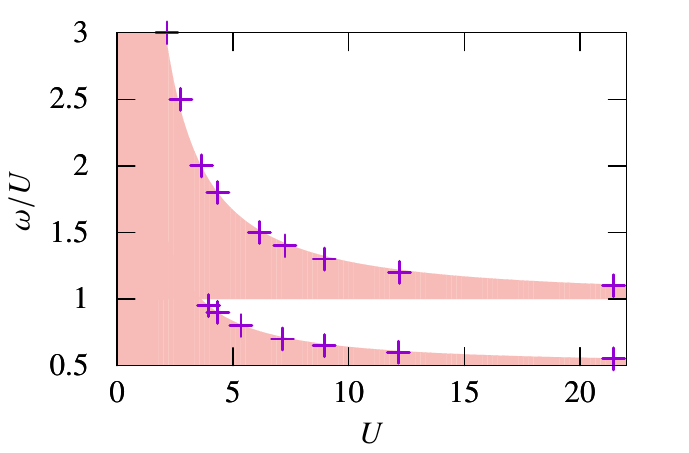}
\caption{Region (shaded) where the level crossing
between the $\hat{D}=0$ and $\hat{D}=1$ sectors emerges, 
thus invalidating the  perturbative treatment, in the periodically driven three-site Hubbard model.
The crosses represent the numerically determined boundaries, 
which can be approximated as $2\omega-U=5U^{-1/4}$ and $\omega-U=5U^{-1/4}$ as indicated by the shading.
\label{fig:docc-boundary}}
\end{centering}
\end{figure}

Since the jumps in the coupling constants make qualitative behavior
in a small-$A$ region significantly deviated from the perturbative results,
we can regard it as a criterion for a breakdown of the expansion.
The boundaries thus obtained are displayed against $U$ and $\omega/U$ in Fig.~\ref{fig:docc-boundary}.

Let us now examine the accuracy of the conventional high-frequency expansion 
for the chiral coupling $J_\chi$ when $\omega/U$ is varied.
Let us first look at Fig.~\ref{fig:repulsive-params}(d), 
where the conventional high-frequency expansion
is applicable to the Hubbard Hamiltonian Eq.~(\ref{eq:td-hubbard}) ($U\ll\omega$), 
yielding the Hubbard model with complex hopping amplitudes.
We can see that this case is also described within the present formalism,
as the result of the high-frequency expansion recovered
when the denominators in 
Eqs.~(\ref{eq:eff-heisenberg}),~(\ref{eq:eff-chiral-1}), and (\ref{eq:eff-chiral-2})
are expanded in $U/\omega$.  
This contrasts with 
Figs.~\ref{fig:repulsive-params}(a)-\ref{fig:repulsive-params}(c), 
where the conventional high-frequency expansion becomes 
inaccurate or even breaks down.  
Interestingly, such regions have chiral coupling that is 
significantly enhanced compared with the case of Fig.~\ref{fig:repulsive-params}(d).

While the chiral coupling (normalized by $1/U^3$) is enhanced when the frequency is lowered
[note the different vertical scales in Figs.~\ref{fig:repulsive-params}(a)-\ref{fig:repulsive-params}(d)], 
the convergence of the numerical result to the perturbative expression is
achieved for smaller $U$ in the high-frequency case. 
Whether the original value (not normalized by $U$) is enhanced or not should hence be examined 
in terms of the ratio, $|J_\chi|/J$, of original values.  
We have retained the sign of $J$ to distinguish between ferromagnetic and antiferromagnetic cases.

\begin{figure*}
\begin{centering}
\includegraphics{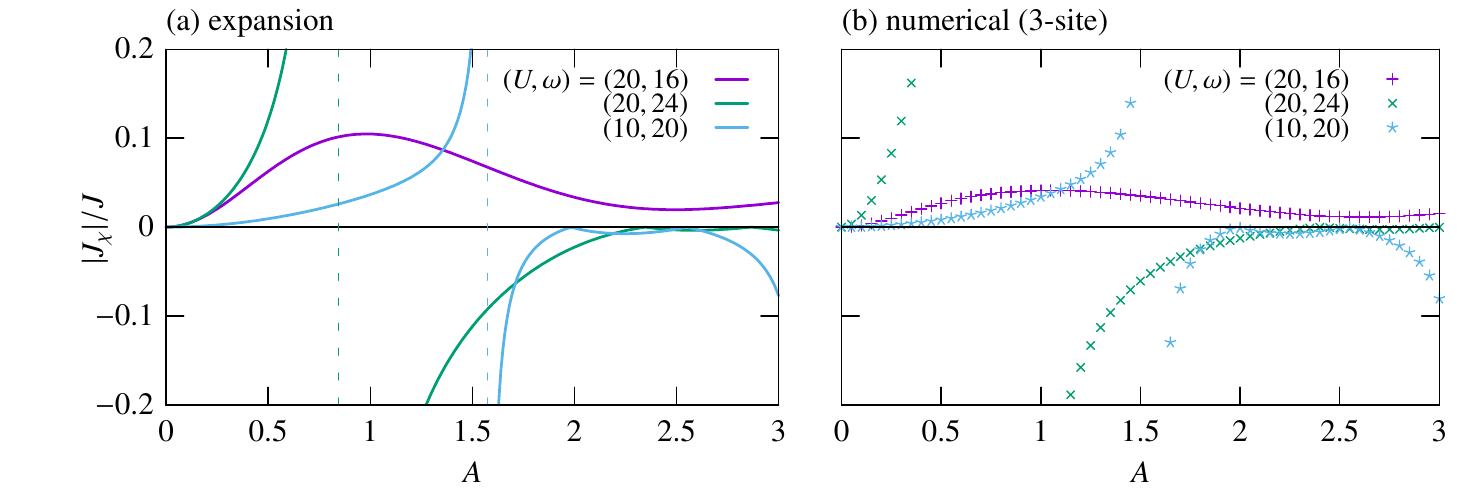}
\caption{The ratio, $|J_{\chi}|/J$, between the chiral coupling and
exchange interaction, plotted against the driving amplitude $A$. 
Vertical dashed lines indicate the values of $A$ for which the exchange interaction $J$ vanishes.
(a) Results for the perturbative expression.
(b) Numerical results for the driven three-site Hubbard model.
\label{fig:repulsive-ratio}}
\end{centering}
\end{figure*}

We plot the ratio against $A$ for various combinations of $(U,\omega)$ in Fig.~\ref{fig:repulsive-ratio}.  
The overall ratio indeed has larger values for the cases corresponding to 
Figs.~\ref{fig:repulsive-params}(b) [for $(U,\omega)=(20,16)$] and ~\ref{fig:repulsive-params}(c) [$(U,\omega)=(20,24)$] 
than that of Fig.~\ref{fig:repulsive-params}(d) [$(U,\omega)=(10,20)$]. 
The ratio diverges when the exchange interaction $J$ vanishes.  
When we have chiral coupling, the possibility arises for the emergence of a chiral spin liquid, which is a topologically nontrivial state. 
This was investigated in a recent study~\cite{Claassen2016}
using the density-matrix renormalization-group method.  
Thus, while the chiral coupling 
$J_{\chi}$ considered here, 
being a higher-order effect, is basically smaller than $J$, 
the coupling can still be significant, exceeding a critical value of $|J_{\chi}|/J$ for the chiral spin liquid, which is estimated to be about 0.16~\cite{Bauer2014}.

On the other hand,
it is imperative to establish a method to detect such nontrivial chiral orders. 
As we shall propose in the next section,
we show that the coupling between scalar chirality and
a circularly polarized laser can be invoked 
to probe the spin chirality.

Before closing this section, let us remark on the heating effects 
implicitly neglected in the present analysis. 
In general, periodically driven macroscopic systems heat up, to 
an infinite temperature in the long-time limit. 
The quasienergy operator describes such infinite-temperature states 
in the thermodynamic limit~\cite{DAlessio2014,Lazarides2014},
which cannot obviously be described within the $\hat{D}=0$ sector 
due to doublon excitations. 
The heating effect can be seen through the breakdown of the strong-coupling expansion 
when higher-order terms are considered.
We have generally a factor of $(mU-n\omega)^{-1}$ in the expansion with 
$0\le m\le \text{min}[M,L]/2$ for the $M$th-order terms in $L$-site systems. 
This factor can become considerably large for arbitrary $U$ and $\omega$~\footnote{
If we define $\delta_k$ through a recurrence relation, $\delta_0=\protect\lceil U/\omega\protect\rceil\omega-U$, 
$\delta_{k+1}=|\protect\lceil\omega/\delta_k-1/2\protect\rceil\delta_k-\omega|$, 
it satisfies $0\le\delta_{k+1}\le\delta_k/2$ 
and has a form $\delta_k=m_kU-n_k\omega$ with integers $m_k\neq0,n_k$.}
as $M,L$ are increased.
This will lead to a breakdown of the expansion, 
which implies that the eigenstate cannot be described within the $\hat{D}=0$ sector 
as in ordinary degenerate perturbation. 

However, even if the expansion is invalidated due to an $M$th-order term, 
one can still truncate the expansion at $(M-1)$th order. 
Then the remaining $\mathcal{O}(\lambda^M)$ terms describe 
transitions toward the $\hat{D}\neq0$ sectors, 
which represent the heating in terms of doublon excitations.
We expect that such $M$ can be made large for appropriately chosen values of $U,\omega$ 
for the heating to be slow enough. 
The heating within the $\hat{D}=0$ sector also exsists, 
due to divergent higher-order terms in the high-frequency expansion Eq.~(\ref{eq:hfe}), which is 
known to be slow in high-frequency driving~\cite{Mori2016,Kuwahara2016}.

\section{Lasers as a probe for spin chiralities}
The strong-coupling expansion has revealed that 
scalar spin chirality couples with circularly polarized laser.
Conversely, this implies that the existence of spin chirality 
in Mott insulators should be fingerprinted in the optical response for 
circularly polarized lasers.
In other words, a Mott insulator should exhibit \textit{circular dichroism}
if it possesses spin chirality.

To examine this, one might intuitively expect that the dielectric function of the present system 
can be obtained by differentiating the effective Hamiltonian $\hat{F}_\text{spin}$ 
with respect to the driving amplitude. 
Whether this holds true is not trivial in general nonequilibrium situations, 
since the effective Hamiltonian does not represent the total energy of the system but the energy up to modulo $\omega$, 
and we may also have to consider nonequilibrium quasienergy distributions. 
However, in the present case, we can justify this in the linear-response regime as follows:

Let us evaluate the dielectric function (tensor) $\varepsilon_{\mu\nu}(\omega)$ with the Kubo formula,
which reads~\cite{Bulaevskii2008}
\begin{equation}
\varepsilon_{\mu\nu}(\omega)=
\delta_{\mu\nu}+\sum_{\alpha\alpha^{\prime}}
\frac{e^{-\beta \epsilon_{\alpha}}-e^{-\beta \epsilon_{\alpha^{\prime}}}}{2Z}
\frac{\langle\Psi_\alpha|\hat{P}^{\mu}|\Psi_{\alpha^{\prime}}\rangle
\langle\Psi_{\alpha^{\prime}}|\hat{P}^{\nu}|\Psi_\alpha\rangle}
{\omega+\epsilon_{\alpha}-\epsilon_{\alpha^{\prime}}+i0^+}.
\label{eq:kubo-formula}
\end{equation}
Here, $|\Psi_\alpha\rangle$ denotes an eigenstate of the undriven system
with an eigenenergy $\epsilon_\alpha$, $\beta$ is the inverse temperature, 
$Z=\sum_\alpha e^{-\beta \epsilon_\alpha}$ 
the partition function, 
$0^+$ a positive infinitesimal, and 
$\hat{P}^\mu$ the $\mu$-component of 
the polarization operator $\hat{\bm{P}}=\sum_{i}\hat{n}_{i}\bm{R}_{i}$.

To relate this expression with spin correlations,
it is convenient to rewrite the expression to obtain
\begin{align}
\varepsilon_{\mu\nu}(\omega)&=
\delta_{\mu\nu}-\sum_{\alpha}\frac{e^{-\beta\epsilon_{\alpha}}}{2Z\omega^{2}}
\langle\Psi_\alpha|[\hat{P}^{\mu},[\hat{P}^{\nu},\hat{H}_{\text{u}}]]|\Psi_\alpha\rangle
\nonumber\\
&-\sum_{\alpha\alpha^{\prime}}\frac{e^{-\beta\epsilon_{\alpha}}}{2Z\omega^{2}}
\frac{\langle\Psi_\alpha|[\hat{P}^{\mu},\hat{H}_{\text{u}}]|\Psi_{\alpha^{\prime}}\rangle
\langle\Psi_{\alpha^{\prime}}|[\hat{P}^{\nu},\hat{H}_{\text{u}}]|\Psi_\alpha\rangle}
{\omega+\epsilon_{\alpha}-\epsilon_{\alpha^{\prime}}+i0^{+}}
\nonumber\\
&+\sum_{\alpha\alpha^{\prime}}\frac{e^{-\beta\epsilon_{\alpha}}}{2Z\omega^{2}}
\frac{\langle\Psi_\alpha|[\hat{P}^{\nu},\hat{H}_{\text{u}}]|\Psi_{\alpha^{\prime}}\rangle
\langle\Psi_{\alpha^{\prime}}|[\hat{P}^{\mu},\hat{H}_{\text{u}}]|\Psi_\alpha\rangle}
{\omega-\epsilon_{\alpha}+\epsilon_{\alpha^{\prime}}+i0^{+}},
\end{align}
with $\hat{H}_\text{u}$ being the Hubbard Hamiltonian for the undriven case.
Here again we consider the strong-coupling expansion for the undriven Hamiltonian, 
where the transformation is generated by $\mathcal{S}_0\equiv\mathcal{S}(t;A=0)$.
When $e^{-\beta U}\ll1$, we can restrict the range of $\alpha$ summation to the $\hat{D}=0$ sector.
By expanding the denominators
with $\epsilon_\alpha-\epsilon_{\alpha^\prime}$ for $\alpha^\prime$ in the $\hat{D}=0$ sector, and
with $\epsilon_\alpha-\epsilon_{\alpha^\prime}-U$ for $\alpha^\prime$ in the $\hat{D}=1$ sector,
we obtain
\begin{multline}
\begin{pmatrix}\varepsilon_{xx}(\omega) & \varepsilon_{xy}(\omega)\\ \varepsilon_{yx}(\omega) & \varepsilon_{yy}(\omega) \end{pmatrix}\\
=\begin{pmatrix}1 & 0\\ 0 & 1 \end{pmatrix}
+\sum_{ij}\dfrac{2|t_{ij}|^{2}}{U(U^{2}-\omega^{2})}
\left(\left\langle \hat{\bm{S}}_{i}\cdot\hat{\bm{S}}_{j}\right\rangle _{\text{th}}-\dfrac{1}{4}\right)
\begin{pmatrix}x_{ij}^2 & x_{ij}y_{ij}\\ x_{ij}y_{ij} & y_{ij}^2 \end{pmatrix}\\
-\sum_{ijk}\dfrac{4|t_{ij}|^{2}|t_{jk}|^{2}\omega(7U^{2}-3\omega^{2})}{U^{2}(U^{2}-\omega^{2})^{3}}
\mathcal{A}_{ijk}
\left\langle (\hat{\bm{S}}_{i}\times\hat{\bm{S}}_{j})\cdot\hat{\bm{S}}_{k}\right\rangle _{\text{th}}
\begin{pmatrix}0 & -i\\
i & 0
\end{pmatrix},\label{eq:dielectric}
\end{multline}
where $\bm{R}_{ij}=(x_{ij},y_{ij})$, $\mathcal{A}_{ijk}\equiv(\bm{R}_{ik}\times\bm{R}_{jk})_{z}/2=-\mathcal{A}_{jik}$
is the area of a triangle enclosed by sites $i,j,k$, and
$\langle\hat{O}\rangle_{\text{th}}=Z^{-1}\sum_{\alpha}e^{-\beta \epsilon_{\alpha}}
\langle \Phi_\alpha|\hat{O}|\Phi_\alpha\rangle$
with $|\Phi_\alpha\rangle=\hat{P}_0e^{i\hat{\mathcal{S}}_0}|\Psi_\alpha\rangle$.
This is indeed related to the effective static Hamiltonian through an expected formula,
\begin{equation}
\varepsilon_{\mu\nu}(\omega)=\delta_{\mu\nu}+2\left.\dfrac{\partial^{2}\langle\hat{F}_{\text{spin}}\rangle_\text{th}}{\partial E_{\mu}^{\ast}\partial E_{\nu}}\right|_{\bm{E}=0}.
\end{equation}

\begin{figure}
\begin{centering}
\includegraphics{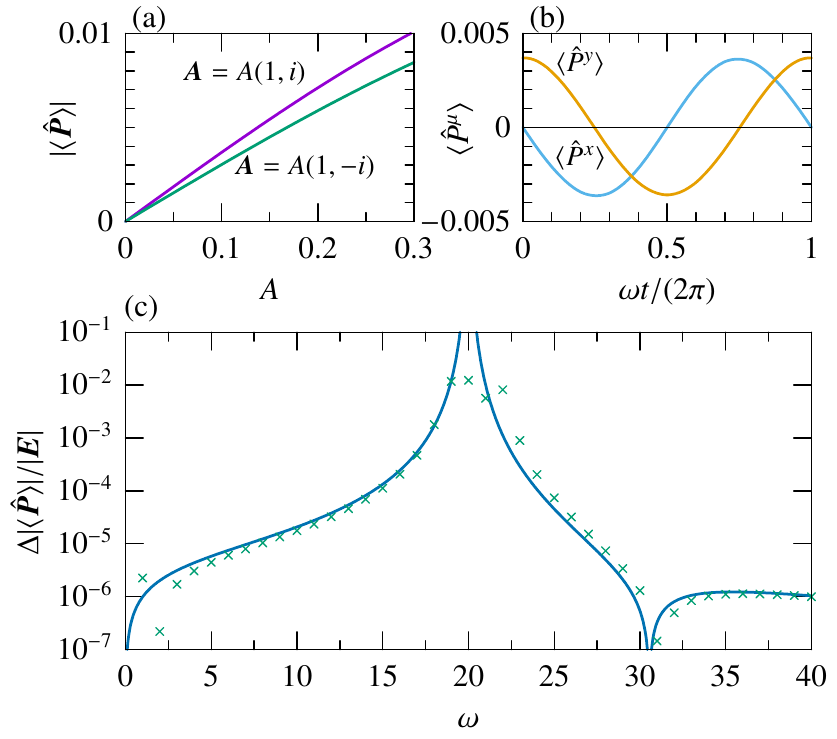}
\caption{Polarization of the ground state for the three-site Hubbard model with an additional chiral term,
$\hat{H}(t)=\hat{H}_\text{Hub}(t)+\gamma(\hat{\bm{S}}_{1}\times\hat{\bm{S}}_{2})\cdot\hat{\bm{S}}_{3}$,
with $t_{ij}=1$, $U=20$, $\gamma=0.01$.
(a) Magnitude of the polarization against $A$ for $\omega=16$ 
for $\bm{A}=A(1,i)$ (a right-circularly-polarized light) or 
$\bm{A}=A(1,-i)$ (left-circularly-polarized). 
(b) Temporal evolution of the polarization $\langle\hat{\bm{P}}\rangle$ 
for $\bm{A}=A(1,i)$, $A=0.1$, $\omega=16$.
(c) Difference, $\Delta|\langle\hat{\bm{P}}\rangle|$, 
in the amplitude between left- and right-circularly-polarized lights, 
normalized by the electric field amplitude.
Crosses represent numerical results, while the solid curve the perturbative expression, 
$|9\omega(7U^2-3\omega^2)/( U^2(U^2 - \omega^2)^3)|$.
\label{fig:polarization}}
\end{centering}
\end{figure}

As Eq.~(\ref{eq:dielectric}) manifestly indicates,
nonzero imaginary offdiagonal parts emerge in the dielectric function
in the presence of the scalar chirality order.
Thus we have the emergence of a circular dichroism, i.e.,
a difference in the reflectivity between left- and 
right-circularly-polarized light. This is actually illustrated in Fig.~\ref{fig:polarization}(a), 
where we consider a three-site Hubbard model with a chiral term,
\begin{equation}
\hat{H}(t)=\hat{H}_\text{Hub}(t)+\gamma(\hat{\bm{S}}_{1}\times\hat{\bm{S}}_{2})\cdot\hat{\bm{S}}_{3},
\end{equation}
which simulates the last term in Eq.~(\ref{eq:dielectric})
to obtain the expectation value of the polarization operator $\hat{\bm{P}}$
for the eigenstate corresponding to the ground state of $\hat{F}_\text{spin}$.
While the direction of the induced polarization rotates as shown in Fig.~\ref{fig:polarization}(b), 
its amplitude is time-independent and exhibits the circular dichroism 
for a right-circularly-polarized light [$\bm{A}=A(1,i)$] or a left-circularly-polarized one [$\bm{A}=A(1,-i)$], 
as seen in Fig.~\ref{fig:polarization}(a).
The difference, $\Delta|\langle\hat{\bm{P}}\rangle|$, 
in the amplitude between left and right circular polarizations 
agrees qualitatively with the predicted value
from Eq.~(\ref{eq:dielectric}), as is evident in Fig.~\ref{fig:polarization}(c).
Namely, we can conclude that the scalar chirality order can be probed
by circularly polarized lasers in the Mott insulators.

Even in the absence of scalar chirality, 
we can point out another possibility for using the circularly polarized laser as a probe
if we exploit cooperative effects when other magnetic orders coexist.
One example is vector spin chirality order. The scalar chirality term would then act as a Zeeman term
if there is a vector spin chirality $\langle\bm{S}_i\times\bm{S}_j\rangle$ 
[see Fig.~\ref{fig:chirality}(b)]
in a mean-field sense~\footnote{
We note that the inverse of this phenomenon may be regarded as 
the twist-exchange interaction proposed in Ref.~\cite{Secchi2013}. 
Namely, the scalar chirality term acts as a Dzyaloshinskii-Moriya interaction 
in the presence of a collinear magnetic order as 
$(\hat{\bm{S}}_{i}\times\hat{\bm{S}}_{j})\cdot\protect\langle\hat{\bm{S}}_{k}\protect\rangle$.}.
Namely, the presence of uniform vector chirality should result in 
laser-induced magnetization, i.e., the inverse Faraday effect.  
Here we have neglected the in-plane 
magnetic components of the oscillating laser field. 
The magnetic component is shown to act as a Zeeman field 
along $z$-axis in the Floquet effective Hamiltonian, 
as shown in Refs.~\cite{Takayoshi2014,Takayoshi2014-2}.
While these contributions are expected to be distinguished by, e.g.,
a frequency or temperature dependence, this should be elaborated upon in future works.

\section{Summary}

In this study, we have investigated a topological feature 
induced by the circularly polarized laser in strongly correlated electron systems.  
We have formulated the strong-coupling expansion for the periodically driven Hubbard model 
to reveal that the scalar spin chirality term emerges 
when the system is driven by a circularly polarized laser.
We have found that the induced scalar chirality should be significant
when the driving field becomes comparable to the driving frequency,
where the conventional high-frequency expansion fails.  
We have further shown that the obtained formula for chiral coupling 
conversely implies a circular dichroism, in the linear-response regime, 
in Mott insulators  when scalar chirality order exists. 

These results indicate novel future directions: 
Now that the present study shows that 
the scalar spin chiral system is not just a toy model 
but can be generated when periodically driven, 
we can question how the emergent scalar chirality will affect spin systems 
with various kinds of lattice structures or interactions.  
The chiral spin liquid phase, e.g. in a Kagom\'e antiferromagnet, is a representative case, as studied in 
Refs.~\cite{He2011,Bauer2014,Hickey2016,Claassen2016}.
Application to known materials such as 
herbertsmithite ZnCu$_{3}$(OH)$_{6}$Cl$_{2}$~\cite{Norman2016}
by combining with first-principles calculations will
be an important direction.   
A related compound, GaCu$_{3}$(OH)$_{6}$Cl$_{2}$, is estimated to
have $U/t\sim17$-$23$~\cite{Mazin2014},
which is suitable in the context of the present discussion.  
Typical orders of magnitude for spin interactions with, say, $A=1$ and $\omega=4$~eV, amount to $J\sim85$~meV and $J_\chi\sim8$~meV.
PdCrO$_2$~\cite{Takatsu2014} is another candidate for application, since it is an antiferromagnet reported to have a scalar spin chirality, hence it is expected to exhibit the circular dichroism proposed above.
Applying the present formalism to the SU($N$) Hubbard model is another interesting path,
which can be realized in cold-atom systems with, e.g., Ytterbium atoms having SU(6)~\cite{Taie2012}.  

\begin{acknowledgments}
T.O. gratefully acknowledges C.~Batista for an illuminating discussion and for explaining the possibility of 
inducing scalar chirality by applying a circularly polarized laser to a Mott insulator in the high-frequency limit. 
We wish to thank Y.-C.~He, Y.~Fuji, M.~Sato, and S.~Takayoshi for fruitful discussions. 
H.A. was supported by ImPACT Program 
of Council for Science, Technology and Innovation, Cabinet Office, 
Government of Japan (Grant No. 2015-PM12-05-01) and MEXT KAKENHI Grant No. JP25107005, while 
S.K. was supported by the Advanced leading graduate course for photon science (ALPS).
\end{acknowledgments}

\bibliography{reference}

\end{document}